\documentclass[reqno]{amsart}
\usepackage[german,english]{babel}  
\usepackage[latin1]{inputenc} 
\usepackage[final]{epsfig}
\usepackage{graphicx}
\usepackage{amssymb}
\usepackage{times} 
\numberwithin{equation}{section}
\newtheorem{theorem}{Theorem}[section]
\newtheorem{proposition}[theorem]{Proposition}
\newtheorem{lemma}[theorem]{Lemma}

\theoremstyle{definition}
\newtheorem{remark}[theorem]{Remark}

\def\supp{\mathop{\hbox{supp}}}
\def\tr{\mathop{\hbox{tr}}}
\def\infspec{\mathop{\hbox{$\mathrm{inf}\,\mathrm{spec}$}}}
\newcommand{\R}{\mathbb{R}}
\newcommand{\Z}{\mathbb{Z}}

\newcommand{\hl}[1]{\frac{#1}{2}}
\newcommand{\po}{\psi_0}
\newcommand{\pbo}{\overline{\psi}_0}
\newcommand{\Vb}{\overline{U}}
\newcommand{\ob}{}
\makeatletter
\def\@tocpagenum#1{\hss{\mdseries #1}}
\def\@tocwrite#1{\@xp\@tocwriteb\csname toc#1\endcsname{#1}}
\def\@tocwriteb#1#2#3{%
  \begingroup
    \def\@tocline##1##2##3##4##5##6{%
      \ifnum##1>\c@tocdepth
      \else \sbox\z@{##5\let\indentlabel\@tochangmeasure##6}\fi}%
    \csname l@#2\endcsname{#1{\csname#2name\endcsname}{\@secnumber}{}}%
  \endgroup
  \addcontentsline{toc}{#2}%
    {\protect#1{\csname#2name\endcsname}{\@secnumber}{#3}}}
\def\l@section{\@tocline{1}{0pt}{0.5pc}{}{}}
\renewcommand{\tocsection}[3]{%
  \indentlabel{\@ifnotempty{#2}{\ignorespaces#1 #2.\quad}}#3}
\def\l@subsection{\@tocline{2}{0pt}{2.5pc}{5pc}{}}

\def\l@subsubsection{\@tocline{3}{0pt}{1pc}{7pc}{}}

\def\l@part{\@tocline{-1}{12pt plus2pt}{0pt}{}{\bfseries}}

\def\l@chapter{\@tocline{0}{8pt plus1pt}{0pt}{}{}}

\makeatother
\begin{document}
\bibliographystyle{alpha}

\title[Random Surface Potentials]{Anderson Localization and Lifshits Tails for \\ Random Surface
Potentials}

\author{Werner Kirsch}
\address{Institut f\"ur Mathematik,
          Ruhr-Universit\"at Bochum, D--44780 Bochum, Germany}%
\email{werner.kirsch@mathphys.ruhr-uni-bochum.de}
\author{Simone Warzel}
\address{Jadwin Hall, Princeton University, NJ 08544, USA.
{\rm On leave from:} Institut f\"ur Theoretische Physik,
                       Universit\"at
                       Erlangen-N\"urnberg, Staudtstra{\ss}e 7, D--91058
                       Erlangen, Germany}%
\email{swarzel@princeton.edu {\rm or:} simone.warzel@physik.uni-erlangen.de}

\begin{abstract}
We consider Schr\"odinger operators on $L^2(\R^d)$ with a random
potential concentrated near the surface
$\R^{d_1}\times\{0\}\subset\R^d$. We prove that the integrated density
of states of such operators exhibits Lifshits tails near the bottom of the spectrum.
From this and the multiscale analysis by Boutet de Monvel and Stollmann [Arch. Math. {\bf 80} (2003) 87]
we infer Anderson localization (pure point spectrum and dynamical localization) for low energies.
Our proof of Lifshits tail relies on spectral properties of Schr\"odinger operators with partially periodic potentials.
In particular, we show that the lowest energy band of such operators is parabolic.
\end{abstract}

\maketitle
\tableofcontents
\section{Introduction}
\subsection{Model}
We consider random Schrödinger operators
\begin{equation}\label{eq:schop}
    H(V\ob):= - \Delta + V\ob
\end{equation}
on the Hilbert space $L^2(\R^d)$ of complex-valued, square-integrable functions on $ \R^d $ with $ d \geq 2 $.
These operators are supposed to
model non-interacting electrons in a (possibly imperfect) $ d $-dimensional crystal with additional
random impurities on the  $ d_2 $-dimensional surface
(or: interface) $ \R^{d_1}\times\{0\}\subset\R^d = \R^{d_1} \times  \R^{d_2}$.
Accordingly, the potential consists of three parts
\begin{equation}\label{eq:defV3}
    V\ob := U_{\mathrm b} + V_{\mathrm b}\ob + V_{\mathrm s}\ob.
\end{equation}
The first part is supposed to model the perfect crystal.
Our assumptions on this non-random part $ U_{\mathrm b} : \R^d \to \R $ of the bulk potential are:
\medskip
\begin{itemize}
\item[{\bf B1}] \begin{itemize}
    \item[1.] $ U_{\mathrm b} $ is periodic with respect to translations of the (sub)lattice $ \mathbb{Z}^{d_1} $:
    $$
    U_{\mathrm b}(x_1+i, x_2) = U_{\mathrm b}(x_1, x_2)
    $$
    for all $ x_1 \in \R^{d_1} $, $ x_2 \in \R^{d_2} $ and all $ i \in \mathbb{Z}^{d_1} $.
    \medskip
    \item[2.]
    $ U_{\mathrm b}\in \mathcal{K}(\R^d) \, \cap \, L^2_{\mathrm{loc}}(\R^d) $.
    \end{itemize}
\end{itemize}
\medskip
For the definition and properties of the Kato class $ \mathcal{K}(\R^d) $, see \cite{Sim82}.
Since B1.2 guarantees that $ H(U_{\mathrm b} ) = - \Delta + U_{\mathrm b} $
is self-adjoint and lower bounded on $ L^2(\R^d) $, the bottom of its spectrum
can be set to zero by a suitable shift in the energy:
\medskip
\begin{itemize}
\item[{\bf~~}]  \begin{itemize}
    \item[3.] $ \infspec H(U_{\mathrm b}) = 0 \quad $ [wlog] \end{itemize}
\end{itemize}
\medskip
As $ U_{\mathrm b} $ is only required to be $ \mathbb{Z}^{d_1} $-periodic, it models not only situations for which
the surface is embedded in in a single crystal but also those for which $ \R^{d_1}\times\{0\} $ acts as an interface
between two different crystals. We may even take $ U_{\mathrm b} $ very large (but bounded) on one side of the interface, such that this
side becomes almost impenetrable for electrons.\\

Both parts of the crystal may (or may not)
contain impurities giving rise to a non-negative random bulk potential, which is
defined on some complete probability space $ (\Omega_{\mathrm b}, \mathcal{A}_{\mathrm b}, \mathbb{P}_{\mathrm b} ) $. Its
realizations are denoted by $  V_{\mathrm b}: \R^d \to [0, \infty[ $ and we will suppose throughout:
\medskip
\begin{itemize}
\item[{\bf B2}] \begin{itemize}
    \item[1.]   $ V_{\mathrm b}\ob  $ is ergodic with respect to translations of the (sub)lattice $ \mathbb{Z}^{d_1} $.
        \medskip
    \item[2.] $ 0\leq V_{\mathrm b}\ob \in L^2_{\mathrm{unif}}(\R^d)  $.
     \medskip
    \item[3.]
    There exists $\kappa_{\mathrm b }>0 $ such that for all $ \varepsilon > 0 $ small enough:
    $$
    \mathbb{P}_{\mathrm b}\left\{  \frac{1}{| \Lambda |} \int_{\Lambda}   V\ob_{\mathrm b}(x)\, dx  < \varepsilon \right\}
    \geq \varepsilon^{\kappa_{\mathrm b } | \Lambda |}
    $$
    for all
    $ \Lambda := \big[ -\frac{L}{2} , \frac{L}{2} ]^{d_1} \times \big[ - \frac{c}{2} \ln L , \frac{c}{2}\ln L  \big]^{d_2}
     \subset \R^d $
    with large enough volume
    $ | \Lambda | := L^{d_1} \big( c \ln L \big)^{d_2} $ and all $ c > 0 $.
    \end{itemize}
\end{itemize}
\medskip
Here the third assumption basically ensures that the probability of $   V\ob_{\mathrm b} $
being tiny on an arbitrarily large set around the interface
is positive. Examples of random potentials fullfilling this assumption are many positive alloy-type random potentials
(cf.\ \cite{Kir89,CaLa90,PaFi92}). We also note that $V_{\mathrm b}$ is allowed to vanish identically.\\

The main emphasis in this paper lies on the presence of a
random surface potential, which is defined on some complete probability
space $ (\Omega_{\mathrm s}, \mathcal{A}_{\mathrm s}, \mathbb{P}_{\mathrm s} ) $.
We will choose its realizations
$ V_{\mathrm s}: \R^d \to ]-\infty,0] $ to be of alloy type
\begin{equation} \label{def:pot}
V_{\mathrm s}\ob(x):=\sum_{i\in\mathbb{Z}^{d_1}} q_i\ob  \; f(x_1-i,x_2)
\end{equation}
where we write $ x = (x_1, x_2) \in \R^{d_1} \times  \R^{d_2} $.
The couplings $\{q_i\}_{i \in \mathbb{Z}^{d_1}} $ are independent, identically distributed random variables with common
distribution $P_0$ and $f: \R^d \to [0, \infty[$ is called single-site potential.
Moreover we assume the $\{q_i\}$ to be independent of $V_{\mathrm b}$.
Throughout this paper we impose the following assumptions:
\medskip
\begin{itemize}
\item[{\bf S1}] \begin{itemize}
    \item[1.] $\supp P_0$ is compact and contained in $]-\infty,0[$,
    it is not concentrated in a single point and if $q_{\mathrm{min}}:=\inf \supp P_0$
then $q_{\mathrm{min}}<0$.
        \medskip
    \item[2.]
    There is some  $\kappa_{\mathrm s }>0$ such that for all $ \varepsilon > 0 $ small enough:
     $$ P_0([q_{\mathrm{min}}, q_{\mathrm{min}}+\varepsilon])\geq \varepsilon^{\kappa_{\mathrm s }}.$$
    \end{itemize}
\medskip
\item[\bf{ S2}] \begin{itemize}
    \item[1.] $f$ is non-negative, positive on a non-empty open set.
    \medskip
    \item[2.]
    $f\in\ell^1(L^p(\mathbb{R}^d))$ with $p\ge 2$ and $p>d$
        \end{itemize}
\end{itemize}
\medskip
The local regularity assumption 2. on $ f $ can be relaxed to $p\ge \max(2,\frac{d}{2})$ for a substantial part
of this paper. However, for technical reasons we need the stronger condition close to the boundary
of regions where we have to impose boundary conditions.

\medskip
In particular, S1 and S2 together with
\medskip
\begin{itemize}
\item[\bf{S3}] \begin{itemize} \item[] $\inf_{x_1\in \mathbb{R}^{d_1}} U_{\mathrm s}(x_1, x_2) \rightarrow 0$ as
$|x_2| \to \infty$. \end{itemize}
\end{itemize}
\medskip
ensure that the partially periodic potential $U_{\mathrm s}: \R^d \to \R $ given by
\begin{equation}\label{eq:sper}
     U_{\mathrm s}(x) := q_{\mathrm{min}} \sum_{i\in\mathbb{Z}^{d_1}}f(x_1-i,x_2)
\end{equation}
is uniformly locally $ p $-integrable, $U_{\mathrm s} \in L^p_{\mathrm{unif}}(\mathbb{R}^d) \subset \mathcal{K}(\R^d) $
    with $ p $ as in S2.\\

Under the above assumptions the random Schr\"odinger operator $ H(V\ob)$ is almost surely essentially self-adjoint
on $ C^\infty_0(\mathbb{R}^d) $, the space of arbitrarily often differentiable functions with compact support (cf.\ \cite{KirMar83a}).
Moreover, $ \mathbb{Z}^{d_1} $-ergodicity of $ V $
on the product measure space $ (\Omega_{\mathrm b} \times \Omega_{\mathrm s} ,
\mathcal{A}_{\mathrm b} \otimes \mathcal{A}_{\mathrm s},  \mathbb{P}  ) $ with $ \mathbb{P} := \mathbb{P}_{\mathrm b} \otimes \mathbb{P}_{\mathrm s} $,
guarantees the validity of (cf. \cite{KirMar82,EKSS90})
\begin{proposition}
    Under assumptions B1--B2 and S1--S3
     the spectrum of $ H(V\ob) $ is almost surely non-random.
    The same applies to the pure point, the singular continuous and the absolutely continuous part of $ {\mathrm{spec}} \,H(V\ob) $.
\end{proposition}
From B2 and S1/2 it follows that $ \infspec H(V) \geq \infspec H_{\mathrm{per}} =: E_0 $, where we introduce the $ \mathbb{Z}^{d_1} $-periodic
background operator
\begin{equation}\label{eq:background}
    H_{\mathrm{per}} := -\Delta + U_{\mathrm b} + U_{\mathrm s}
\end{equation}
on $ L^2(\R^d) $. Using techniques developed for bulk random potentials \cite{Kir89,CaLa90,PaFi92},
it is not hard to show that the above lower bound is actually an equality. However, the following proposition can also be viewed
as a corrollary to Theorem~\ref{thm:Lif} below.
\begin{proposition}
Under assumptions B1--B2 and S1--S3 we have $  \infspec H(V) = E_0 $.
\end{proposition}
In the present paper we will always assume:
\medskip
\begin{itemize}\item[{\bf S4}]  \begin{itemize} \item[]
$E_0 = \infspec H_{\mathrm{per}}  < 0  \quad \big[ = \infspec H(U_{\mathrm b})  \big] $ \end{itemize}
\end{itemize}
\medskip
It can be shown that $E_0<0$ if $q_{\mathrm{min}}<q(d_2)$ where $q(d_2)=0$ for $d_2=1$ and $d_2=2$, but negative in higher
dimension (cf.\ \cite[Thm.~XIII.11/12]{ReSi4}).
Basically, S4 gurantees that all eigenstate $ \psi_E $ of $ H(V) $ corresponding to negative eigenvalues
$ E < 0 $ are concentrated near the internal surface,
\begin{equation}
    \sup_{x_1 \in \R^{d_1}} | \psi_E (x_1,x_2) | \leq C e^{- \gamma |x_2| }
\end{equation}
for some constants $ \gamma $, $ C > 0 $
(see Theorem~\ref{prop:exp} below). Moreover, these surface states are energetically
separated from the spectrum of $ H(U_{\mathrm b} + V_{\mathrm b} ) $,
which occurs above zero.
However, we would like to warn the reader that, in contrast to what the symbols suggests,
even $ H(U_{\mathrm b}) $ may have (generalized) eigenstates
which are concentrated near $ \R^{d_1}\times\{0\} $ (for a discussion see \cite{DaSi78}, and also \cite{EKSS90}).

\subsection{Main results}
Under assumptions B1--B2 and S1--S4 we first prove the existence of the
integrated density of surface states (IDSS) for negative energies, that is, below the spectrum of the bulk operator (cf.\ S4).
For its definition we set
\begin{equation}\label{eq:SL}
    S_L:=\Lambda_L\times \R^{d_2} \quad \mbox{with} \quad \Lambda_L:= \left[-\frac{L}{2},\frac{L}{2}\right]^{d_1}
    \quad \mbox{and} \quad L \in \mathbb{N},
\end{equation}
a strip around the $x_2$-direction. At our convenience and when it does not cause confusion,
we will also write $ H:= H(V\ob) $ and
drop the dependence on the potential.
Accordingly, we denote by $H^X_{S_L}(V\ob) =: H^X_{S_L} $ the
operator (\ref{eq:schop}) restricted to $L^2(S_L)$ with $ X $-boundary conditions
at $\partial S_L$, where $ X = D $ or $ X = N $ stands for
Dirichlet respectively Neumann boundary conditions. Its eigenvalue-counting function
\begin{equation}\label{eq:2}
N\big(H^X_{S_L} ,E\big) :=\# \Big\{n\in \mathbb{N}_0 \, \mid \, E_n\big(H^X_{S_L} \big)\leq E\Big\}
\end{equation}
is called the reduced-volume IDSS.
Here we introduce the notation $E_0(A) \leq E_1 (A) \leq \dots $ for the eigenvalues of a self-adjoint operator $A$ in
increasing order and counted according to multiplicity and set $E_n(A)=\infspec_{\mathrm{ess}}A $ if $A$ has at
most $n-1$ eigenvalues below its essential spectrum. Thanks to B1 and S3 and the Weyl theorem \cite[Thm.~XIII.14]{ReSi4}, the
essential spectrum of $H^N_{S_L}(V) \leq H^D_{S_L}(V) $ is contained in
$[0,\infty[$, so that (\ref{eq:2}) is well defined for negative energies $E<0$.

The following theorem allows us to define the IDSS $ N(E) := N^X(E) $ for all $ E < 0 $
as the (unique left-continuous) infinite-volume limit of their reduced-volume
counterparts.
\begin{theorem}\label{thm:IDSS}
Under assumptions B1--B2 and S1--S4 the limit
\begin{equation}\label{eq:3}
N{}^X(E) :=\lim_{L \rightarrow \infty}\frac{N\big(H^X_{S_L}\!(V),E\big)}{L^{d_1}}, \quad E < 0,
\end{equation}
exists and is almost surely non-random for both $ X = D $ and $ X = N $.
Moreover, $ N^D(E) = N^N(E) $ for all $ E < 0 $ except for countable many.
\end{theorem}

The IDSS was first rigorously examined in \cite{EKSS88,EKSS90}
and further investigated in
\cite{JaMoPa98,Ch99,KoSch00,KoSch01,Boe03}.
These authors define $ N $  in slightly different ways and consider the IDSS at all energies by
subtracting the IDSS of the bulk operator.
We will discuss this issue in Section~\ref{sec:2} and show that, below zero, these alternative definitions give the
same quantity as the limit in (\ref{eq:3}) as long as $ f $ has compact support in $x_2$-direction
(cf.\ (\ref{eq:comp}) below; an assumption made in all of the above mentioned works).
\\

The two main purposes of this paper are to prove that $ N(E) $ exhibits Lifshits tails near $ E= E_0 $ and to conclude
Anderson localization therefrom. Our first result concerns the Lifshits tails for rapidly decaying $ f $ in the sense of
\medskip
\begin{itemize}\item[{\bf S5}]
$ f(x_1, x_2) \leq f_0 \, | x_1 |^{-d_1-2} $ for some constant $ f_0 $ and $|x_1|$ large.
\end{itemize}
\medskip
Here the additional local assumption 2 on $ f $ is mainly of technical origin.
\begin{theorem} \label{thm:Lif}
Under assumptions B1--B2 and S1--S5 we have
\begin{equation}\label{eq:Lif}
    \lim_{E\downarrow
    E_0}\frac{\ln|\ln N{}(E)|}{\ln(E-E_0)} =
    -\frac{d_1}{2}.
\end{equation}
\end{theorem}
We can also handle single-site potentials $ f$ which decay
slower than $|x|^{-d_1-2}$. In fact, replacing the decay requirement in S5 by
\medskip
\begin{itemize}
        \item[{\bf S5'}] \begin{itemize}
    \item[] There exist constants $f_u $,$ f_0>0$ and a non-empty open Borel set
    $ F_2 \subset \mathbb{R}^{d_2} $ such that
        $ f_u \, | x_1 |^{-\alpha} 1_{F_2}(x_2)
    \le f(x_1, x_2) \leq f_0 \, |x_1|^{-\alpha} $
        for some $d_1 < \alpha \leq d_1 + 2$, large $| x_1|$ and all $ x_2 \in \mathbb{R}^{d_2}$.\newline
        [Here $ 1_F $ denotes the characteristic function of a (Borel) set  $ F $.]
    \end{itemize}\end{itemize}
\medskip
we obtain
\begin{theorem} \label{thm:Lif2}
    Under assumptions B1--B2 and S1--S5' we have
    \begin{equation}\label{eq:Lif2}
     \lim_{E\downarrow
    E_0}\frac{\ln|\ln N{}(E)|}{\ln(E-E_0)}=-\frac{d_1}{\alpha-d_1}.
\end{equation}
\end{theorem}
The proof of both theorems
basically follows old strategies developed in \cite{KirSim86,Mez87,KiWa} for bulk tails.
At its core, however, lie some new results on partially periodic potentials, which are presented in Section~\ref{Sec:per}.
These results are interesting in their own and their proof relies on a method of seperable comparison potentials developed in
Subsection~\ref{Subsec:gap}.

We finally remark that an analysis of the Lifshits tails for discrete surface operators
was given in \cite{KiKl}. This paper also deals with the case $E_0=0$,
a case we can not handle here.\\

Localization of surface states by (alloy-type) random surface potentials is discussed in detail in \cite{BdMS03},
which has been the main motivation of the present paper.
In fact, in case $ U_{\mathrm b} = V_{\mathrm b} = 0 $ and under the assumption that $P_0$ is H\"older continuous
\medskip
    \begin{itemize}
        \item[{\bf S6}]\begin{itemize}
    \item[] There exist constants $C$, $\mu > 0$
        such that $P_0([a,b]) \le C\,(b-a)^\mu$ for all $a<b$.
    \end{itemize}\end{itemize}
\medskip
and the additional assumption that $P_0([q_{\mathrm{min}},q_{\mathrm{min}}+\varepsilon])$ decays
sufficiently fast at $q_{\mathrm{min}} =\inf \supp P_0$, Boutet de Monvel and
Stollmann \cite{BdMS03} prove spectral and dynamical localization.
Theorem~\ref{thm:Lif} allows us to obtain their result for the present model without this
additional decay assumption on $P_0$.
\begin{theorem}\label{thm:loc}
    Suppose assumptions B1--B2 and S1--S6 hold. Then
    \begin{itemize}
    \item[a)] there exists an energy $E_1 > E_0  $
    such that almost surely $H(V) $ has pure point spectrum in $[E_0,E_1 ]$ with exponentially
    decaying eigenfunctions.
    \item[b)] there exists an energy $E_1>E_0$
    such that in $I=[E_0,E_1]$ we have:
    \begin{equation}
     \mathbb{E}\,\left(\,\sup_{t>0} \left\| |x|^p\,  e^{itH(V) } P_I(H(V)) \, 1_K\right\|\right)\;<\;\infty
     \end{equation}
    for any compact set $K\subset\R^d$.
    {\rm[}Here $ \mathbb{E}$ denotes expectation with respect to $ \mathbb{P}=\mathbb{P}_{\mathrm b} \otimes
    \mathbb{P}_{\mathrm s} $ and $ P_I(H) $ stands
    for the spectral projection of $ H $ associated with $ I $.{\rm]}
    \end{itemize}
\end{theorem}
Let us finally remark that there is a vast literature on the spectral structure of Schr\"odinger operators with
random surface potentials, which mostly deals with the discrete case (see \cite{JaMoPa98,JaMo99,ChSa00,JaLa00,JaLa01}
and references therein). The only works other than \cite{BdMS03,BKS04} (and the present paper)
investigating continuum models are \cite{HuKi00,BdMSS04}. In case $ U_{\mathrm b} = V_{\mathrm b} = 0 $
they show the presence of absolutely continuous
(bulk) spectrum at non-negative energies \cite[Thm.~4.3]{HuKi00} and
the absence of absolutely continuous spectrum at negative energies in case $ d_1 = 1 $ \cite[Thm.~4.1]{BdMSS04}.

\section{Existence of the integrated density of surface states}\label{sec:2}
The IDSS was first
introduced in \cite{EKSS88,EKSS90} as a distribution (of order at most $3$).
In particular in case $ U_{\mathrm b} = V_{\mathrm b} = 0 $, the authors proved that the limit
\begin{equation}\label{def:dossEKSS}
\nu{}(\varphi):=\lim_{L\rightarrow\infty}\frac{1}{L^{d_1}}\:
    \tr\left[1_{C_L}\left[\varphi(H(V_{\mathrm s}\ob))-\varphi(-\Delta)\right]1_{C_L}\right]
\end{equation}
\medskip
of expanding cubes $C_L=[-\frac{L}{2},\frac{L}{2}]^d$ defines a non-random linear functional $ \nu{} $
on $\varphi \in C_0^\infty(\R)$.
Actually \cite{EKSS90} considered the interface between two different random
potentials and \cite{KoSch00} remarked that the method in \cite{EKSS90} can be used
for surface potentials as well. The renormalization term
$\varphi(-\Delta) $ in (\ref{def:dossEKSS}) is needed to counterbalance the
first term which diverges as soon as $\varphi$ has support inside the spectrum of
$-\Delta$. This term is not needed below $0=\infspec (-\Delta)$, in fact, it vanishes there.
Hence $\nu{}(\varphi) \geq 0 $ for $\varphi\geq0$ having support
below zero, so that $\nu{}$ restricted to Borel subsets of $]-\infty,0[ $ is a non-negative measure.

Kostrykin and Schader \cite{KoSch00,KoSch01} defined the IDSS in a slightly different way. They look at the
operator $H\big(V_{\mathrm s} 1_{C_L}\big) = - \Delta + V_{\mathrm s} 1_{C_L} $
with the
potential cut off outside the cube $C_L$ and proved that
\begin{equation}\label{def:dossKS}
\lim_{L \to \infty} \frac{1}{L^{d_1}}\;\tr\left[\varphi\left(H\big(V_{\mathrm s} 1_{C_L}\big)\right)-\varphi(-\Delta)\right]
\end{equation}
exists and agrees with $\nu{}(\varphi)$ as in (\ref{def:dossEKSS}). For this Kostrykin and Schrader
assumed that $V_{\mathrm s}\ob$ is of the form (\ref{def:pot}) with $f$ compactly supported in $x_2$-direction, that is,
\begin{equation}\label{eq:comp}
    \supp f \subset \R^{d_1} \times \big[-L,L]^{d_2}
\end{equation}
for $L$ large enough.
They also proved regularity properties of $\nu{}$
inside $\mathrm{spec}(-\Delta) = [0 , \infty[ $ which improved the results from \cite{EKSS90} considerably.

It is not hard to see that that the (non-negative) distribution function $ \nu{}(\,]-\infty,E]) $ for $ E < 0 $ corresponding to
Kostrykin-Schrader's definition (\ref{def:dossKS}) and hence to (\ref{def:dossEKSS})
coincides with the IDSS defined through the
limit in (\ref{eq:3}),
\begin{equation}
  N(E) = N^X(E) = \nu{}(\,]-\infty,E])
\end{equation}
for all $ E < 0 $ except countably many.
In fact, by Neumann-Dirichlet-bracketing (\cite[Prop.~3 on p.~269]{ReSi4} or \cite{KirMar82b}) we
have $ H_{S_L}^N( V_{\mathrm s}\ob ) \oplus H_{\complement S_L}^N(0)  \leq H( V_{\mathrm s}\ob 1_{S_L})
\leq H_{S_L}^D( V_{\mathrm s}\ob ) \oplus H_{\complement S_L}^D(0) $, which gives
\begin{equation}\label{est:Hschl}
N\big(H_{S_L}^D(V_{\mathrm s}\ob), E\big)\leq N\big(H(V_{\mathrm s}\ob 1_{S_L}),E\big)\leq N\big(H_{S_L}^N(V_{\mathrm s}\ob), E\big),
\end{equation}
because $ N\big( H_{\complement S_L}^N(0), E \big) = 0 $ for $ E<0$.

\begin{remark}
Definition (\ref{def:dossEKSS}) as well as
definition (\ref{def:dossKS}) cannot be used if $  0 \in \supp \varphi $ and in case $f$ decays in
$x_2$-direction slower than $|x_2|^{-2}$. Instead of $ V_{\mathrm s}\ob $ let us
take a non-random potential $W \leq 0 $ which depends only on $ x_ 2 $ and decays slowly.
Then the operator $H(W) $ separates into the free
Laplacian in $x_1$-direction and the operator $- \Delta +W$ in
$x_2$-direction. The latter has infinitely many
eigenvalues below zero (cf.\ \cite[Thm. XIII.82]{ReSi4}).
Hence the $\varphi(-\Delta)$-term in (\ref{def:dossEKSS}) (or in (\ref{def:dossKS}))
has no chance to smooth out the singularity at $ 0 $ originated by
those eigenvalues.
\end{remark}

\subsection{Nonisotropic exponential decay of eigenfunctions}
An important ingredient in our proof of Theorem~\ref{thm:IDSS} will be the exponential decay of eigenfunctions along
the $ x_2 $-direction, which correspond
to eigenvalues below $ \infspec H(U_{\mathrm b}) = 0 $ and in particular below the essential spectrum of  $H_{S_L}^X\!(V\ob)$.
We remark that similar steps were used in proofs in \cite{Boe03,BoeKir}.
\begin{theorem} \label{prop:exp}
Suppose assumptions B1.2--3, B2.2, S1.2 and S2--S4 hold and let $ \eta < 0 $.
There are constants $C$, $\gamma > 0 $ such that for both $ X = D $ and $ X = N $, every $ L \in [ 1 , \infty]$
and every $ L^2(S_L) $-normalized eigenfunction
$\psi_E$ of $H_{S_L}^X\!(V\ob) $ corresponding to an eigenvalue $E\leq\eta$ one has
\begin{equation} \label{eq:expon}
    \sup_{{x_1} \in [-\hl{L},\hl{L}]^{d_1}} | \psi_E(x_1,x_2) | \leq C \, e^{-\gamma |x_2|}
\end{equation}
for  $| x_2 |$ large.
\end{theorem}
\begin{proof}
Since $\psi_E$ is an eigenfunction, we have $ \psi_E=\exp\left[-t\big(H_{S_L}^X\!(V\ob)-E\big)\right]\psi_E $
for all $ t \geq 0 $.
Using the Feynman-Kac-formula (cf.\ \cite{Sim79}) we write the semigroup as an
integral over Brownian paths $\beta: [0, \infty[ \to  S_L $, which start at $ x \in S_L $ for
$t=0$ and have either absorbing boundary
conditions (in case $X=D$) or reflecting boundary conditions (in case $X=N$, see \cite[Thm.~6.3.12]{BraRob81})
at $\partial S_L$. Denoting the corresponding Wiener measure by
$p^X_x $, we have
\begin{equation}
  \left| \psi_E(x)\right| \leq \int \exp\left[\int_0^t\big(E - V\ob(\beta(s))\big)\,ds\right]
        \,\left|\psi_E\big(\beta(t)\big)\right| \; p^X_x(d\beta).
\end{equation}
To estimate the integral from above we first observe that B2.2, S1.2 and S2 with (\ref{eq:defV3}) implies
$ V\ob \geq U_{\mathrm b} + V_{\mathrm s}\ob $ and that
\begin{equation}\label{eq:nachunten}
V_{\mathrm s}\ob(x_1,x_2)\geq q_{\mathrm{min}} \sum_{i\in \Z^{d_1}}f(x_1-i,x_2) =
U_{\mathrm s}(x_1,x_2) \geq \inf_{x_1\in \mathbb{R}^{d_2}} U_{\mathrm s}(x_1,x_2).
\end{equation}
By S3 this term goes to zero as $|\,x_2|\rightarrow \infty$, so that
$E-V_{\mathrm s}\ob(x_1,x_2)\leq\frac{\eta}{2}<0$ for $|x_2| $ large enough. We therefore split
the $\beta$-integration into an integration over
$   \Omega_1 := \big\{\beta \,\mid \, \sup_{0\leq s\leq t} \, |\beta_2(s)-x_2|<\frac{|x_2|}{2} \big\} $
and its complement. Taking $|x_2| $ large, we thus have
\begin{multline}
|\psi_E(x)| \leq e^{t \frac{\eta}{2}} \int_{\Omega_1} \exp\left[ - \int_0^t \, U_{\mathrm b}(\beta(s))\,ds\right]\,
    \left|\psi_E\big(\beta(t)\big)\right|\; p^X_x(d\beta) \\
    +  \int_{\complement \Omega_1} \exp\left[ \int_0^t \big(E - U(\beta(s))\big)\,ds\right]\,
    \left|\psi_E\big(\beta(t)\big)\right|\; p^X_x(d\beta) \label{eq:2terme}
\end{multline}
where we introduced the abbreviation $ U :=  U_{\mathrm b} + U_{\mathrm s} $ in the last term.

Dropping the restriction to $ \Omega_1 $,
the first Wiener integral in (\ref{eq:2terme}) defines the ultracontractive
semigroup $ \exp\big[-t H_{S_L}^X\!(U_{\mathrm b})\big] $ from $ L^2(S_L) $ to $ L^\infty(S_L) $. In fact, the integral is estimated by
\begin{equation}\label{eq:2unendl}
 \sup_{x\in S_L}\left( e^{-t H_{S_L}^X\!(U_{\mathrm b})} \big|\psi_E\big|\right)(x)
       \leq \big\| e^{-t H_{S_L}^X\!(U_{\mathrm b})}\big\|_{2,\infty} \, \big\|\psi_E \big\|_2  \leq C.
\end{equation}
Here we have introduced $ \| \cdot \|_{p,q} $ for the norm of a bounded operator from $ L^p(S_L) $
to $ L^q(S_L) $. The second is valid for $ t \geq 1 $ and follows from B1.2--3 and Lemma~\ref{lemma:Kato} below.

Using the Cauchy-Schwarz inequality and $ E \leq 0 $, the second Wiener integral in (\ref{eq:2terme}) is bounded from above by
\begin{equation}\label{eq:CS}
 p^X_x\big(\complement\Omega_1\big)^{\frac{1}{2}} \left(\int
    \exp\left[-2\int_0^t U(\beta(s))\,ds\right] \, \left|\psi_E\big(\beta(t)\big)\right|^2\; p^X_x(d\beta)\right)^{\frac{1}{2}}.
\end{equation}
The Wiener integral in (\ref{eq:CS})
defines an ultracontractive semigroup $ \exp\big[-t H_{S_L}^X\!(2 U)\big] $, which is bounded from $ L^1(S_L) $
to $ L^\infty(S_L) $
according to Lemma~\ref{lemma:Kato} below. In fact,
the integral is bounded from above by\footnote{
Note that the constants in (\ref{eq:2unendl}) and (\ref{eq:1unedl}) differ; we will nevertheless subsequently use the
same symbol $ C $ for occurring constants.}
\begin{equation}\label{eq:1unedl}
    \sup_{x\in S_L} \Big(e^{-t H_{S_L}^X\!(2U)} \, \big| \psi_E |^2 \Big)(x)
     \leq \big\| e^{-t H_{S_L}^X\!(2U)}\big\|_{1,\infty} \big\|\psi_E \big\|_2^2  \leq C \, \exp\big[t \upsilon\big]
\end{equation}
where $ \upsilon := | \infspec H(2U ) | < \infty $.
We finally note that the first factor in (\ref{eq:CS}) is
exponentially small in $|x_2|$. Since $\complement  \Omega_1 $ only involves $ \beta_2 $, its $ p^X_x $-measure equals
its Wiener measure $ p_x $ on Brownian path $ \beta $, which start at $ x \in \R^d $ and wind through in all of $ \R^d $.
Using Levy's maximal inequality
\cite[Eq.~(7.6')]{Sim79} (for the last $ d_2 $ components $ \beta_2: [0,t] \to \mathbb{R}^{d_2} $ of Brownian motion)
we have
\begin{align}
p^X_x\big(\complement \Omega_1\big)&=
p_0\!\left(\sup_{0
\leq s \leq t} |\beta_2 (s)| \geq \frac{|x_2|}{2}\right) \notag \\
& \leq  2\,p_0\!\left(|\beta_2(t)| \geq \frac{|x_2|}{2}\right) \leq \;4\, e^{-\frac{|x_2|^2}{32t}}
\end{align}
Gathering terms and choosing $t =  \max\{ |x_2| / 32 \sqrt{\upsilon}, 1\} $, we obtain the desired result.
\end{proof}
The above proof made use of the following lemma, which in case $ X = D $ is well-known in the theory of Schr\"odinger semigroups
\cite{Sim82}  (see also \cite[Eq.~(2.40)]{BHL00}). As we could note find the result for $ X = N $, we include it for the reader's
convenience.
\begin{lemma}\label{lemma:Kato}
    Let $ W: \R^d \to \R $ with $ W^+ \in \mathcal{K}_{\mathrm{loc}}(\R^d) $ and $ W^- \in \mathcal{K}(\R^d) $, where
    $ W^\pm(x) = \sup\{\pm W(x), 0 \}$.
    Then the semigroup $ \exp\!\big[ - t H_{S_L}^X\!(W) \big] $ is
    bounded from $ L^q(S_L) $ to $ L^\infty(S_L) $, $ q \in \{1,2\} $, and there is some constant $ C $ such that
    \begin{equation}\label{eq:Kato}
        \big\| e^{-t H_{S_L}^X\!(W)} \big\|_{q, \infty} \leq C \, \exp\!\big[  - t \, \infspec H(W) \big]
    \end{equation}
    for all $ L \in[ 1, \infty] $, all $ t \geq 1 $ and both $ X= D $ and $ X = N $.
\end{lemma}
\begin{proof}
The semigroup property and duality implies that
\begin{equation}
    \big\| e^{-t H_{S_L}^X\!(W)} \big\|_{1, \infty} \leq
    \big\| e^{-\frac{t}{2} H_{S_L}^X\!(W)} \big\|_{2, \infty} \big\| e^{-\frac{t}{2} H_{S_L}^X\!(W)} \big\|_{1, 2}
    =   \big\| e^{-\frac{t}{2} H_{S_L}^X\!(W)} \big\|_{2, \infty}^2
\end{equation}
It therefore remains to investigate $ \exp\big[ - t H_{S_L}^X\!(W) \big] $ from $ L^2(S_L) $ to $ L^\infty(S_L) $. Using the semigroup
property again we find
\begin{align}
    \big\| e^{-tH_{S_L}^X\!(W)} \big\|_{2, \infty} & \leq
    \big\| e^{-\tau H_{S_L}^X\!(W)} \big\|_{2, \infty} \big\| e^{- (t -\tau) H_{S_L}^X\!(W)} \big\|_{2, 2} \notag \\
    & \leq \big\| e^{- \tau H_{S_L}^X\!(W)} \big\|_{2, \infty} \, \exp\!\big[ - (t - \tau) \, \infspec H_{S_L}^X\!(W) \big]
    \label{eq:tah}
\end{align}
for every $ 0 < \tau < t $, which gives the exponential factor in (\ref{eq:Kato}) since $ \infspec H_{S_L}^X\!(W) \geq \infspec H(W) $.
In case $ X = D $ a finite, $L $-independent upper bound on the first factor in (\ref{eq:tah}) is provided by
$ \big\| e^{- \tau H(W)} \big\|_{2, \infty} \geq \big\| e^{- \tau H_{S_L}^D\!(W)} \big\|_{2, \infty} $, since the Dirichlet semigroup is
increasing in the domain.
In case $ X = N $ we first note that the integral kernel $ e^{\tau \Delta_{S_L}^N}(x,y) $ of the semigroup generated by
the Neumann Laplacian can be explicitely computed using the known \cite[p.~266]{ReSi4}
eigenfunctions of $ \Delta_{\Lambda_L}^N $. It is bounded according to
\begin{equation}\label{eq:kern}
 e^{\tau \Delta_{S_L}^N}(x,y) \leq (4 \pi \tau)^{-\frac{d}{2}} \left( 1 + (4 \pi \tau)^{\frac{1}{2}}\right)^{d_1} \,
    \exp\left[- \frac{|x_2 - y_2|^2}{4 \tau}\right]
\end{equation}
for all $  L \in[ 1, \infty] $.
Since $ W^- $ is infinitesimally form bounded with respect to the Neumann Laplacian,
for every $ \varepsilon > 0 $ there is an $ L$-independent constant $ C_\varepsilon > 0 $
such that $\big\langle \psi , H_{S_L}^N\!(W) \, \psi \big\rangle \geq \big\langle \psi , H_{S_L}^N\!(-W^-) \, \psi \big\rangle \geq
    (1-\varepsilon) \big\| \nabla \psi \big\|_2^2 - C_\varepsilon \, \| \psi \|_2^2 $
for all $ \psi \in W^{1,2}(S_L) $. The assertion $ \big\| e^{- H_{S_L}^D\!(W)} \big\|_{2, \infty} \leq  C $,
then follows from (\ref{eq:kern}), the above lower bound and \cite[Thm.~2.4.6 and Cor.~2.4.3]{Dav89},
which relates the ultracontractivity of
semigroups to the form of their generators.
\end{proof}

\subsection{Proof of Theorem~\ref{thm:IDSS}}
Our first step will be the proof of the existence and non-randomness of the infinite-volume limits in Theorem~\ref{thm:IDSS}.
\begin{proposition}\label{lemmma:ex}
The limit in (\ref{eq:3}) exists and is almost surely non-random for both for $X=D$ and $X=N$.
\end{proposition}
\begin{proof} The stochastic process $ N\big(H^X_{S_L}\!(V\ob),E\big)$ indexed by cubes
$\Lambda_L\subset\R^{d_1}$ (cf.\ (\ref{eq:SL})) is superadditive for $X=D$ and
subadditive for $X=N$. We may therefor apply the Akcoglu-Krengel ergodic
theorem \cite{AkKr81,Kre85} (see also \cite{KirMar82b}), since $ V $ is $ \R^{d_1} $-ergodic and
$\mathbb{E}\left[ N\big(H^X_{S_1}(V\ob),E\big)\right] $ is finite, in fact uniformly bounded.
\end{proof}

To actually prove that $ N^D = N^N $ requires a little more work. We even introduce additional
quantities which could also be used to define the IDSS (cf.\ Theorem~\ref{thm:rho} below). While we employ them
mainly as auxiliary tools, we believe that they are interesting in themselves.
We set
\begin{equation}\label{eq:defL}
    \Lambda_{L,M} :=\left[-\frac{L}{2},\frac{L}{2}\right]^{d_1}
    \times\left[-\frac{M}{2},\frac{M}{2}\right]^{d_2}
\end{equation}
and let  $ H_{L,M}^{X,Y}(V\ob) =:  H_{L,M}^{X,Y} $ (we again drop the dependence on the potential at our convenience)
be the operator (\ref{eq:schop}) restricted to $ L^2(\Lambda_{L,M})$ where $X$
and $Y$ refer to either Dirichlet or Neumann boundary conditions.
In particular, $H_{L,M}^{X,Y}$ has $X$-boundary conditions on
$\partial\,[\!-\!\frac{L}{2},\frac{L}{2}]^{d_1}\times ]\!-\!\frac{M}{2},\frac{M}{2}[^{d_2}$
and $Y$-boundary conditions on
$]\!-\!\frac{L}{2},\frac{L}{2}[^{d_1}\times\partial\,[\!-\!\frac{M}{2},\frac{M}{2}]^{d_2}$.\\

Our first step is to compare the eigenvalue counting function of $ H_{S_L}^X $ with that of $ H_{L,M}^{X,D} $.
\begin{lemma}\label{thm:ND}
Let $\eta < 0$ and $L \geq 1 $. There exist constants $\alpha$, $ M_0$, $ C> 0$ such that
\begin{equation}\label{neuDir}
N\big(H_{L,M}^{X,D},E\big) \leq N\big(H_{S_L}^X, E\big) \leq N\big(H_{L,M}^{X,D},E+C L^{d_1} e^{-\alpha M}\big)
\end{equation}
for both $X=D$ and $X=N$, all $M \geq M_0$ and all $E \leq \eta$.
\end{lemma}
\begin{proof}
The first inequality follows from Neumann-Dirichlet-bracketing. In fact,
$H_{S_L}^X \leq H_{L,M}^{X,D} \oplus H_{S_L \setminus \Lambda_{L,M}}^{X,D} $
and consequently
\begin{equation}
    N\big(H_{S_L}^X,E\big) \geq N\big(H_{L,M}^{X,D},E\big) + N\big(H_{S_L \setminus \Lambda_{L,M}}^{X,D},E\big)
    \geq N\big(H_{L,M}^{X,D},E\big).
\end{equation}
To prove the second inequality in (\ref{neuDir}) we take a complete set $\psi_{E_0}, \dots, \psi_{E_r} $ of
$ L^2(S_L) $-normalized eigenfunction of $H_{S_L}^X$ with eigenvalues $ E_0 \leq \dots \leq E_r \leq E \leq \eta $
and use them to construct
approximate eigenfunction of  $H_{L,M}^{X,D}$.
For this purpose we choose a smooth characteristic function $\tilde 1_M \in C^\infty(\mathbb{R}^d)$ of the set
$ \widetilde \Lambda_M := \mathbb{R}^{d_1} \times \big[ -\frac{M}{2}, \frac{M}{2}\big]^{d_2} $,
which has the property that $ \tilde 1_M(x) = 0 $  for $x \in \partial \widetilde \Lambda_M  $,
$ \tilde 1_M(x) = 1 $ for $x \in \widetilde \Lambda_{M-1} $ and $ \| \nabla \tilde 1_M \|_\infty $, $ \| \Delta \tilde 1_M \|_\infty < C $
for some constant $ C $.

According to Theorem~\ref{prop:exp} each $ \psi_{E_j} $ is exponentially decaying such that there are constants $ \alpha $, $M_0> 0 $
to ensure
\begin{equation}\label{est:ef1}
\left| \left\langle \tilde 1_M\,\psi_{E_j}, \tilde 1_M\,\psi_{E_k} \right\rangle - \delta_{jk} \right| \leq C L^{d_1} e^{-\alpha M}
\end{equation}
for all $ j$, $k \in \{ 0, 1, \dots, r \} $ and all $ M \geq M_0 $. For the same reason and since $  \tilde 1_M \psi_{E_k} $ complies with
Dirichlet boundary conditions, we also have
\begin{equation}\label{est:ef2}
  \left| \left\langle \tilde 1_M\psi_{E_j}\, , H_{L,M}^{X,D} \; \tilde 1_M \,\psi_{E_k} \right\rangle - E_j \,\delta_{jk} \right|
    \leq C L^{d_1} e^{-\alpha M}.
\end{equation}
In fact, this inequality follows from the eigenvalue equation and the product rule, which yield
\begin{equation}
     H_{L,M}^{X,D} \; \tilde 1_M \,\psi_{E_k} = {E_k}\,  \tilde 1_M \,\psi_{E_k} - \psi_{E_k}
\, \Delta \tilde 1_M - 2 \, \big( \nabla \tilde 1_M \big) \!\cdot \!\big( \nabla \psi_{E_k} \big),
\end{equation}
together with (\ref{est:ef1}) and a local gradient estimate \cite[Lemma~2.6]{CyFr87}.
By choosing $ M_0 $ large enough, the upper bound in (\ref{neuDir}) is a consequence of Lemma~\ref{lem:eig} below.
\end{proof}
The above lemma implies
\begin{theorem}\label{thm:rho}
    For all but countably many $E<0$ and any $\varrho>0$:
    \begin{equation}
    \lim_{L\rightarrow\infty}
    \frac{1}{L^{d_1}}N(H_{L,L^\varrho}^{X,D},E)=N{}^X(E)
    \end{equation}
    for both $ X=D$ and $  X=N$.
\end{theorem}

\begin{proof}
 By Proposition~\ref{lemmma:ex} and the lower bound in (\ref{neuDir}) we have
 \begin{equation}
    N^X(E) = \lim_{L \to \infty} \frac{1}{L^{d_1}} N\big(H_{S_L}^{X},E\big)
    \geq \limsup_{L \to \infty} \frac{1}{L^{d_1}} N\big(H_{L,L^\varrho}^{X,D},E\big).
\end{equation}
Using the upper bound in (\ref{neuDir}), we may further estimate
\begin{align}
 \liminf_{L \to \infty} \frac{1}{L^{d_1}} N\big(H_{L,L^\varrho}^{X,D},E\big) & \geq
 \liminf_{L \to \infty} \frac{1}{L^{d_1}} N\big(H_{S_L}^{X},E - C L^{d_1} e^{-\alpha L^{\varrho}}\big) \notag \\
    & \geq  \liminf_{\varepsilon \downarrow 0}N^X(E-\varepsilon).
\end{align}
The last inequality follows from Proposition~\ref{lemmma:ex} and the fact that $ N\big(H_{S_L}^{D},E \big) $ is non-decreasing in $ E $.
Since $N{}^X$ is nondecreasing, it is continuous at all $E<0$ with
the exception of at most countably many points.
\end{proof}

We are now ready to complete the proof of Theorem~\ref{thm:IDSS}.
\begin{proposition} $N{}^D(E)=N{}^N(E) $
for all $E<0$ except at most countably many.
\end{proposition}
\begin{proof}
We let $ E < 0 $ and use a Laplace transform estimate \cite[Lemma~3.3]{KirMar82b}
\begin{multline}
 \int_{-\infty}^E \!\left[ N\big(H_{L,L^\varrho}^{N,D},E'\big)-N\big(H_{L,L^\varrho}^{D,D},E'\big)  \right] dE'
\leq e^{E} \; \tr\!\left[e^{- H_{L,L^\varrho}^{N,D}}-e^{-H_{L,L^\varrho}^{D,D}}\right] \\
\leq e^{E}\; \left(\tr\,e^{-H_{L,L^\varrho}^{N,D}(q U)}\right)^{\frac{1}{q}}\;
\left(\tr\left[e^{\Delta_{L,L^\varrho}^{N,D}}-e^{\Delta_{L,L^\varrho}^{D,D}}\right]\right)^{\frac{1}{p}}
\label{eq:traces}
\end{multline}
where  $1 \leq p$, $q \leq \infty$ with
$\frac{1}{p}+\frac{1}{q}=1$.
By assumptions B1, S2 and S3 we have $U:= U_{\mathrm b} + U_{\mathrm s} \in L^p_{\mathrm{unif}}(\mathbb{R}^d)$,
such that the first trace involving $ - \Delta + qU $ is bounded by $C L^{d_1}L^{\varrho d_2}$.
By a direct computation \cite[p.~266]{ReSi4} the second trace in (\ref{eq:traces}) is bounded by $C L^{d_1-1}L^{\varrho
d_2}$, so that (\ref{eq:traces}) is actually bounded by a constant times
\begin{equation}
    e^{E}\;  L^{\frac{d_1}{q}} L^{\frac{d_2\,\varrho}{q}}
L^{\frac{d_1-1}{p}} L^{\frac{d_2\,\varrho}{p}} \leq e^{E}
L^{d_1 + \varrho d_2 -\frac{1}{p}}.
\end{equation}
Choosing $\varrho<\frac{1}{p d_2}$ the exponent in the right-hand side is less than
$d_1$. Dividing by $ L^{d_1} $ and taking the limit $ L \to \infty $ shows that the right-hand side of (\ref{eq:traces})
tends to zero. Thanks to positivity, $ L^{-d_1} $ times its integrand therefore tends to zero for almost all $ E' \leq E $. By
Theorem~\ref{thm:rho} this proves the assertion.
\end{proof}

\begin{remark}
We finally remark that the same method of proof also shows that the
infinite-volume limit of $H_{L,L^\varrho}^{X,N}$ (instead of
$H_{L,L^\varrho}^{X,D}$) would again give $N{}^N(E)=N{}^D(E)$.
Moreover, a close look at the arguments used in the proof of Theorem~\ref{prop:exp}
shows that an analogous result holds for the eigenfunctions of $H_{L,M}^{X,Y} $.
More precisely, for $ \eta < 0 $
there exists constants $ M_0$, $ C $, $ \gamma > 0 $ such that for every $ L \geq 1 $, every $ M \geq M_0 $ and every
$ L^2(\Lambda_{L,M}) $-normalized eigenfunction $ \psi_{E}^{M} $ of $ H_{L,M}^{X,Y} $  corresponding to an eigenvalue $ E \leq \eta $
one has
\begin{equation}\sup_{x_1 \in
[-\hl{L},\hl{L}]^{d_1}} |\,\psi_E^M(x_1,x_2)| \leq C e^{-\gamma |x_2|},
\end{equation}
for all $ | x_2 | \leq M /2 $.
\end{remark}

\subsection{Appendix: Variational estimate}
The following lemma has been used in the proof of the upper bound in Lemma~\ref{thm:ND}.
It seems to be folklore in Hilbert space theory. However, as we could not find it in the literature,
we include it for the reader's convenience (see also \cite{Boe03} for a similar result).
\begin{lemma}\label{lem:eig}
Let $ n \in \mathbb{N} $ and $\varphi_1, \dots, \varphi_n \in \mathcal{H} $ be
in the domain $ \mathrm{dom}\, A $ of a self-adjoint operator $ A $, which acts on a (separable) Hilbert space $\mathcal{H}$.
Suppose there are constants $ \alpha_1 \leq \dots \leq \alpha_n \leq \alpha $ such that
\begin{equation}\label{eq:eig}
    |\langle\varphi_i, \varphi_j\rangle - \delta_{i,j}| \leq \varepsilon_1  \quad \mbox{and} \quad
        | \langle\varphi_i, A \varphi_j\rangle - \alpha_j \delta_{i,j} | \leq  \varepsilon_2
\end{equation}
for all $ i $, $j = 1, \dots , n $.
If $ \varepsilon_1 $ is small enough, then $
     N\big(A, \frac{\alpha + \varepsilon_2}{1-\varepsilon_1}\big) \geq n $.
\end{lemma}
\begin{proof}
It is well known (cf.\ \cite[Thm.~6 in Ch.~9]{BirSol87}) that the eigenvalue-counting function is given by
\begin{equation}
     N(A, \alpha') = \sup_{ \mathcal{V} \subset \mathrm{dom} A} \big\{  \dim \mathcal{V} \;  |
    \; \langle \phi , A \phi \rangle \leq \alpha' \langle \phi , \phi \rangle \quad \mbox{for all} \quad
    \phi \in \mathcal{V}  \big\}
\end{equation}
in terms of a supremum of the dimensions of all linear
subspaces $ \mathcal{V} $ in the domain of $ A $.
If $ \varepsilon_1 $ is small enough, then $ \varphi_1, \dots, \varphi_n $ span a subspace $ \mathcal{V}_n $ of dimension $ n $.
Moreover, for every $ \phi \in  \mathcal{V}_n $ there exist (non-unique) coefficients $ c_1, \dots, c_n \in \mathbb{C} $ such that
$ \phi = \sum_{j=1}^n c_j \varphi_j $. Thanks to (\ref{eq:eig}) one has
\begin{equation}
\langle \phi , \phi \rangle
    \geq \sum_{j=1}^n |c_j |^2 - \sum_{j,k=1}^n | c_j | |c_k| \, \left|\langle\varphi_i, \varphi_j\rangle - \delta_{i,j}\right|
    \geq (1- \varepsilon_1) \sum_{j=1}^n |c_j |^2
\end{equation}
and similarly
\begin{equation}
    \langle \phi , A \phi \rangle \leq \sum_{j=1}^n \alpha_j \, |c_j |^2 + \sum_{j,k=1}^n | c_j | |c_k| \,
    \left| \langle \varphi_i , A \varphi_j \rangle - \alpha_j \delta_{i,j} \right| \leq ( \alpha + \varepsilon_2) \sum_{j=1}^n |c_j |^2.
\end{equation}
Setting $ \alpha' = (\alpha + \varepsilon_2)/(1-\varepsilon_1)  $ completes the proof.
\end{proof}
\section{Partially periodic Potentials}\label{Sec:per}
In this section we analyze operators
\begin{equation}\label{eq:persufH}
H_{\mathrm{per}}=-\Delta + U
\end{equation}
on $ L^2(\mathbb{R}^d) $ with a partially periodic potential,
that is, $U : \mathbb{R}^d \to \mathbb{R} $ with the property
\begin{equation}
U(x_1+i,x_2) = U(x_1,x_2)
\end{equation}
for all $i\in\Z^{d_1}$. Throughout this section we assume $U \in \mathcal{K}(\R^d) \cap L^2_{\mathrm{loc}}(\R^d)$, 
which is implied, for example, by assumptions S2,S3.\\
This ensures that $ H(U) $ is essentially self-adjoint on $ C_0^\infty(\R^d) $ (cf.\ \cite{CyFr87}).
Examples of partially periodic potentials are fully periodic ones, ``surface periodic'' potentials as in (\ref{eq:sper}) or, more interestingly, commensurable combinations as in (\ref{eq:background}).
Aspects of the spectral (in particular: scattering) theory of such potentials are studied in \cite{DaSi78}.
\subsection{Basic properties\label{sect:perbas}}
Analogous to the fully periodic case \cite[Ch.~XIII.16]{ReSi4}
we can (partially) Floquet-Bloch decompose $H_{\mathrm{per}}$ into a family of
reduced operators $h_\theta $ indexed by $ \theta\in[-\pi,\pi[^{d_1}$, the (partial) Brillouin zone.
We define $ h_\theta $ as the differential operator (\ref{eq:persufH})
acting on $L^2(S_1) $  with $\theta$-periodic boundary conditions on $ \partial S_1 $.
Denoting by $e_j$, ${j\in\{1,\ldots,d_1\}}$, the canonical unit vectors of the subspace
$ \mathbb{R}^{d_1} \times \{0\} \subset \mathbb{R}^d $, we thus
require for functions in the domain of the Laplacian in (\ref{eq:persufH}):
\begin{equation}\label{eq:domain}
\varphi(x+e_j) = e^{i\,\theta_j}\; \varphi(x)  \qquad
    \frac{\partial \varphi}{\partial x_j}(x+e_j) = e^{i\,\theta_j} \, \frac{\partial \varphi}{\partial x_j}(x)
\end{equation}
if both $x$ and $x+e_j$ belong to $S_1$.
Defining a unitary operator
$ \mathcal{U}: L^2(\R^d) \to (2 \pi)^{-d_1} $\hspace{0pt}$ \int^{\bigoplus}_{[-\pi,\pi[^{d_1}}\! L^2(S_1) \, d\theta  $
by setting
\begin{equation}
    \left(\mathcal{U}\psi\right)_\theta(x):= \sum_{n\in \mathbb{Z}^{d_1}}  e^{-i \theta \cdot n} \, \psi(x_1 + n , x_2)
\end{equation}
for functions $ \psi $ in the
Schwartz space $ \mathcal{S}(\mathbb{R}^d) $ and unitarily extending to all of $ L^2(\R^d) $, we then obtain \cite[Sec.~5]{DaSi78}
(see also \cite[Thm.~XIII.97]{ReSi4} and \cite[Sec.~2]{Ch00} for the discrete case)
\begin{proposition}\label{prop-dirint}
Under assumption P1 the operator $ H_{\mathrm{per}} $ is unitarily equivalent to a direct integral of operators,
\begin{equation}
     \mathcal{U} \, H_{\mathrm{per}} \, \mathcal{U}^{-1} = \int^{\bigoplus}_{[-\pi,\pi[^{d_1}}\! h_\theta\;\frac{d\theta \, }{(2 \pi)^{d_1}}.
\end{equation}
\end{proposition}

The only substantial difference to the fully periodic case (as in \cite[Ch.XIII.16]{ReSi4}) is the fact that the operators
$h_\theta$ have essential spectrum.

\begin{remark}
    Analogous to the fully periodic case one can show that $ h_\theta $ is unitarily equivalent to
    $ h_0 - 2 i \theta \cdot \nabla + | \theta |^2 $ defined on the domain of $ h_0 $ (cf.\ (\ref{eq:domain})).
    Analytic pertubation theory \cite[Thm.~XII.8]{ReSi4} then guarantees that $ t \mapsto E_0(h_{t\theta}) $
    is analytic in a complex neighborhood of $ [0,1] \ni t $ for every $ \theta $.
    In particular, $  E_0(h_{\theta}) = \infspec h_\theta $ is a (simple)
    eigenvalue if and only if $  E_0(h_{0}) $ is a (simple) eigenvalue.
\end{remark}

\begin{proposition}\label{prop-per}
The infinum $E_0$ of the spectrum of $H_{per}$ agrees with the infinum $E_0(h_0)$ of the spectrum of $h_0$.
Moreover, if $E_0(h_0)$ belongs to the discrete spectrum of $h_0$, it is a simple eigenvalue and its eigenfunction
can be chosen to be positive.
\end{proposition}
\begin{proof}
The semigroup $e^{-th_0}$ is positivity improving hence there is a nonzero distributional solution
of the equation
\begin{equation}
 h_0 u=E_0(h_0)u
\end{equation}
on the "torus"\ $\R^d/\Z^{d_1}$, which is everywhere non-negative
by Allegretto-Piepenbrink theory (see \cite[Thm.~C.8.1]{Sim82} and its proof)).
The periodic extension of $u$ gives a distributional solution of $H_{per}u=E_0(h_0)u$
which is everywhere nonnegative. Hence, again by Allegretto-Piepenbrink,
$E_0(h_0)\leq E_0$. The reverse inequality is obvious from Proposition~\ref{prop-dirint}.\\
If $E_0(h_0)$ is an eigenvalue of $h_0$, the nondegeneracy follows from the fact that
$e^{-th_0}$ is positivity improving (see \cite[Thm.~XIII.45(a)]{ReSi4}).
\end{proof}

We found it worthwhile to notice that the band function $ \theta \mapsto E_0(h_\theta) $ corresponding to the lowest eigenvalue
of $ h_\theta $ is parabolic in the sense of
\begin{theorem}\label{thm:para}
Under assumptions P1--P2 The ground-state band of $H_{\mathrm{per}}$ obeys
\begin{equation}\label{eq:para}
C \;|\theta|^{2} \le E_0(h_\theta) - E_0(h_0) \le  |\theta|^{2}
\end{equation}
for some $ C >0 $ and all $ \theta \in [-\pi,\pi[^{d_1} $.
\end{theorem}
This theorem is not needed in the following. It serves as another application of the
method of separable comparison potential introduced in the next subsection.
We give its proof in Subsection~\ref{Subsec:parabolicity}

\subsection{Gap estimate}\label{Subsec:gap}

In this section we prove a lower bound on the gap of $H_{\mathrm{per}}$ on $L^2(S_L)$. It is a key ingredient in our proof of Lifshits
tails in Section~\ref{sec:Lif}, in which we have to restrict the operator $H_{\mathrm{per}}$ to strips $S_L$ and
special Robin boundary conditions, which we dubbed Mezincescu boundary conditions in \cite{KiWa}.

To introduce Mezincescu boundary conditions we additionally assume
$ U \in L_{\mathrm{loc}}^p(S) $ for some neighborhood $ S \subset \mathbb{R}^d $ of $ \partial S_1 $ and some $ p > d $.
This is guaranteed by S2. We note that it is only here where we use that $p > d$ instead of $p > d/2$.
This ensures \cite[Thm.~C.2.4]{Sim82} that the ground state $\psi_0$ of $H_{\mathrm{per}}$
can be chosen not only positive and $\Z^{d_1}$-periodic (cf.\ Proposition~\ref{prop-per}) but also continuously differentiable in
a neighborhood of $ \partial S_L $ for every $ L \in \mathbb{N}$. We set
\begin{equation}\label{def:chi}
\chi(x) := -\frac{1}{\psi_0(x)}\, \nabla_n \, \psi_0(x) \qquad\mbox{ for } x\in\partial S_L,
\end{equation}
where $\nabla_n$ denotes the outer normal derivative at the boundary $\partial\,S_L$ of the strip $S_L$, and
define the operator $H_{\mathrm{per},_L}^\chi $ as the restriction of (\ref{eq:persufH}) to $ L^2(S_L) $ with Mezincescu boundary conditions
\begin{equation}\label{eq:mezbo}
\nabla_n\,\varphi(x)=-\chi(x)\,\varphi(x) \qquad\mbox{ for } x\in\partial S_L
\end{equation}
in the domain of the Laplacian (for details of the definition of $ - \Delta^\chi_{S_L} $
via quadratic forms, see \cite{Mez87} and \cite[Sec.~3.1]{KiWa}).
This choice of boundary conditions ensures that
\begin{equation}\label{eq:infgleich}
    \infspec H_{\mathrm{per},L}^\chi  = \infspec H_{\mathrm{per}} = E_0
\end{equation}
for all $ L \in \mathbb{N} $.
Since (\ref{eq:infgleich}) is in general wrong for the operator $H_{\mathrm{per},L}^N $ with Neumann boundary consditions,
Mezincescu boundary conditions were introduced in \cite{Mez87} to be able to extend the result in \cite{KirSim86}.\\

The result which plays a crucial role in our Lifshits-tail estimate is the following lower bound for the gap of $H_{\mathrm{per},L}^\chi$.
\begin{theorem}\label{thm:gapest}
Under assumptions P1--P3 the lowest and second lowest eigenvalue  of $H_{\mathrm{per},L}^\chi$
obey
\begin{equation}\label{eq:gapest}
E_1\big(H_{\mathrm{per},L}^\chi\big)-E_0\big(H_{\mathrm{per},L}^\chi\big) \ge \frac{C_{\mathrm{per}}}{L^2}
\end{equation}
for some $ C_{\mathrm{per}} > 0 $ (which depends on $ U $) and $ L \in \mathbb{N} $ large enough.
\end{theorem}
\begin{proof}
We will prove this theorem in a number of steps. First, we assume that the partially periodic potential $U $ does not depend on
$x_1$.
\begin{lemma}\label{prop:gapsep}
If $U $ is independent of $x_1$, then {\rm (\ref{eq:gapest})} holds with $ C_{\mathrm{per}} = \pi^2 $.
\end{lemma}
\begin{proof}
Since $ U $ is independent of $ x_1 $, the eigenvalue problem separates.
Moreover, the ground state $\psi_0$ of $H_{\mathrm{per}}$ is independent of $x_1$, so that Mezincescu and Neumann
boundary conditions agree in the present case. The eigenvalues of $H_{\mathrm{per},L}^\chi $ are given by the sum
of the eigenvalues of the Neumann (=Mezincescu) Laplacian $ - \Delta_{\Lambda_L}^N $ on $L^2(\Lambda_L)$
and the negative eigenvalues of $ H_2 := -\Delta_{\mathbb{R}^{d_2}} + U$ on $L^2(\R^{d_2})$.
By a direct computation \cite[p.~266]{ReSi4}
one has $E_0\big(-\Delta_{\Lambda_L}^N\big) =0 $ and $ E_1\big(-\Delta_{\Lambda_L}^N\big)  = \pi^2 L^{-2}$.
Moreover, P2 implies that $ E_0(H_2) = E_0(h_0) $ is an eigenvalue of $ H_2 $ such that
$ E_1( H_2)- E_0( H_2) \ge C $ for some $ C> 0 $.
For large $L$, we thus have
\begin{equation}
    E_0(H_{\mathrm{per},L}^\chi) = E_0\big( H_2 \big)  \quad \mbox{and}\quad
    E_1(H_{\mathrm{per},L}^\chi) = E_0\big( H_2 \big) + \pi^2 L^{-2},
\end{equation}
which gives (\ref{eq:gapest}).
\end{proof}

To extend the above proposition to the general case of Theorem~\ref{thm:gapest}, we proceed as follows.
We let $\psi_0$ be the positive, $L^2(S_1)$-normalized ground state of $H_{\mathrm{per}}$ and set
\begin{equation}
\overline{\psi}_0(x_2) := \int_{\Lambda_1}\;\psi_0(x_1+\xi,x_2)\:d\,\xi.
\end{equation}
Note that $\pbo : \mathbb{R}^d \to ]\, 0 , \infty [ $ does not depend on $x_1$, since $\po$ is $ \mathbb{Z}^{d_1} $-periodic by Proposition~\ref{prop-per}.
Moreover one has
\begin{proposition}\label{prop-barest}
There are positive constants $C_1$, $C_2> 0 $ such that
\begin{equation}\label{eq-barest}
C_1\,\pbo(x_2)\;\le\:\po(x_1,x_2)\:\le\;C_2\:\pbo(x_2)
\end{equation}
for all $x_1\in\R^{d_1}$ and $x_2\in\R^{d_2}$.
\end{proposition}

\begin{proof}
By periodicity we may assume that $x_1\in\Lambda_1$. Since $\po$ is positive and solves the Schr\"odinger equation
$H_{\mathrm{per}} \po  = E_0 \po$, we may apply Harnack's inequality \cite{AiSi82} (see also \cite[Thm.~2.5]{CyFr87} and for explicit constants \cite{HiKa90})  to obtain:
\begin{equation}
C_1\,\po(x_1',x_2)\;\le\:\po(x_1,x_2)\:\le\;C_2\:\po(x_1',x_2)
\end{equation}
for all $x_1,x_1'\in\Lambda_1$. Since the local Kato norms of $U \in \mathcal{K}(\R^d) $ are uniformly bounded, the constants $C_1$, $C_2$
are independent of $x_2$. Hence integration over $x_1' \in \Lambda_1 $ yields (\ref{eq-barest}).
\end{proof}

We may now introduce an averaged potential $ \Vb: \mathbb{R}^d \to \mathbb{R} $  corresponding to $ U $ by
\begin{equation}\label{eq:defVb}
 \Vb(x_2) := \frac{\int_{\Lambda_1} U(x_1, x_2) \, \psi_0(x_1, x_2) \, d x_1}{\int_{\Lambda_1} \psi_0(x_1, x_2) \, d x_1}
\end{equation}
which does not depend on $ x_1 $ by construction.

\begin{lemma}
    Assumptions P1 and P3 hold for $ \Vb $.
\end{lemma}
\begin{proof}
    Both assertions follow from the bound
    $ |\Vb(x_2) \leq \big(C_2/C_1)^2 \int_{\Lambda_1} | U(x_1,x_2) | \, dx_1 $, which results from (\ref{eq:defVb}) and
    Proposition~\ref{prop-barest}.
\end{proof}

By calculating the $d$-dimensional distributional Laplacian of $ \pbo $ and using the fact that $ H_{\mathrm{per}} \, \po = E_0 \, \po $,
we reveal that
\begin{align}
    \Delta \, \pbo(x) & = \int_{\Lambda_1} \Delta \, \psi_0(x_1+\xi,x_2)\:d\,\xi \notag \\
    & = \int_{\Lambda_1} \big[ U(x_1 + \xi,x_2) -E_0 \big] \, \psi_0(x_1+\xi,x_2)\:d\,\xi  \notag \\
    & = \Vb(x_2) \, \pbo(x) - E_0 \, \pbo(x). \label{eq:nablapbo}
\end{align}
Hence $ \overline{H}_{\mathrm{per}} \, \pbo = E_0 \, \pbo $, so that $ \pbo $ is the positive ground state of $ \overline{H}_{\mathrm{per}} := - \Delta + \Vb $ with the same eigenvalue $ E_0 $ as for $ H_{\mathrm{per}} $. Moreover, denoting by $ \overline{h}_0 $ the
operator  $ - \Delta + \Vb $ on $ L^2(S_1) $ with periodic boundary conditions on $\partial S_1 $, the positivity and
$ \mathbb{Z}^{d_1} $-periodicity of $ \pbo $
and (\ref{eq:nablapbo}) imply that $ E_0 $ is the lowest eigenvalue of $ \overline{h}_0 $.
We summarize our findings in
\begin{proposition}
    One has
    $ E_0 = \infspec \overline{H}_{\mathrm{per}}  =  \infspec \overline{h}_0 $. Moreover,  $ \infspec \overline{h}_0 $
    is a (simple) eigenvalue of $ \overline{h}_0 $ with corresponding eigenfunction $ \pbo \in L^2(S_1) $.
\end{proposition}

We may therefore apply Lemma~\ref{prop:gapsep} to obtain
\begin{equation}\label{eq:gapbar}
    E_1\Big(\overline{H}_{\mathrm{per},L}^{\, \overline{\chi}}\Big) - E_0\Big(\overline{H}_{\mathrm{per},L}^{\, \overline{\chi}}\Big)\ge \pi^2\,L^{-2}
\end{equation}
where $ \overline{\chi} $ is defined as in (\ref{def:chi}) with $ \po $ replaced by $ \pbo $.
The proof of Theorem~\ref{thm:gapest} is now completed by using Proposition~\ref{prop-barest} and the comparison theorem
for gaps of Schr\"odinger operators \cite[Thm.~1.4]{KirSim87}.
For its application we note that $ H_{\mathrm{per},L}^\chi $ as well as $ \overline{H}_{\mathrm{per},L}^{\, \overline{\chi}} $ may be
realized as Dirichlet forms. In particular, one has
\begin{multline}
    E_1\big(H_{\mathrm{per},L}^\chi\big)-E_0\big(H_{\mathrm{per},L}^\chi\big) =
    \inf\left\{ \big\| \nabla \varphi \big\|_{\po,L}^2 \, \mid \,  \big\| \varphi \big\|_{\po,L}^2 := \right. \\ \left.
    \! \int_{S_L} \! |\varphi(x) |^2 \psi_0(x)^2 dx = 1   \quad\mbox{and}\quad
     \int_{S_L}  \varphi(x)  \psi_0(x)^2 dx = 0 \right\}
\end{multline}
where the infimum ranges over the domain of the Dirichlet form  $ \| \nabla \varphi \|_{\po,L}^2 $ on the weighted Hilbert space
$ L^2(S_L , \psi_0^2 dx) $, and similarly for $ \overline{H}_{\mathrm{per},L}^{\, \overline{\chi}} $.
Proposition~\ref{prop-barest} and \cite[Thm.~1.4]{KirSim87} therefore yield
\begin{equation}
E_1\big(H_{\mathrm{per},L}^\chi\big)-E_0\big(H_{\mathrm{per},L}^\chi\big) \geq \left(\frac{C_1}{C_2}\right)^2
\left[ E_1\Big(\overline{H}_{\mathrm{per},L}^{\, \overline{\chi}}\Big) - E_0\Big(\overline{H}_{\mathrm{per},L}^{\, \overline{\chi}}\Big)
\right]
\end{equation}
which together with (\ref{eq:gapbar}) finshes the proof of Theorem~\ref{thm:gapest}.
\end{proof}

\subsection{Appendix: Parabolicity of the ground-state band}\label{Subsec:parabolicity}

Our goal in this supplementary subsection is to prove Theorem~\ref{thm:para}.
Let $ \psi_0 $ be the positive, $ L^2(S_1) $-normalized ground state function of $ h_{0} $ and introduce the
subspace
\begin{multline}
    \mathcal{H}_\theta  :=
    \left\{ \varphi \in L^2(S_1,\psi_0^2\, dx) \,: \, \mbox{For all $j=1,\ldots,d_1$:} \right. \\
    \left.  \varphi(x + e_j ) = e^{i\theta_j} \, \varphi(x) \quad \mbox{if both $ x  $ and $ x + e_j $ belong to $ S_1 $}\right\}
\end{multline}
of the weighted Hilbert space $ L^2(S_1,\psi_0^2 \, dx) $, which is equipped with the norm
\begin{equation}
     \| \varphi \|_\theta := \left(\int_{S_1} | \varphi(x) |^2 \, \psi_0(x)^2 \, dx \right)^{1/2}.
\end{equation}
Similarly as in the fully periodic case \cite[Eq.~(2.3)]{KirSim87} one can show that
\begin{equation}\label{eq:Dirichlet}
     E_0(h_\theta) - E_0(h_0) =
    \inf\left\{ \|  \nabla \varphi \|_\theta^2 \, \mid \, \| \varphi \|_\theta = 1 \right\}.
\end{equation}
Here the infimum is taken over all functions in domain
$ \{ \varphi \in \mathcal{H}_\theta  \, \mid \, \nabla \varphi \in \mathcal{H}_\theta \} $ of the
Dirichlet form in the right-hand side.
As far as the upper bound  in (\ref{eq:para}) is concerned,
we use $ \varphi(x) = \exp(i \theta \cdot x_1) $ as a variational function in (\ref{eq:Dirichlet}) to obtain
\begin{equation}
     E_0(h_\theta) - E_0(h_0) \leq |\theta|^2.
\end{equation}
For a proof of the lower bound in (\ref{eq:para}) we note that the gap
of the reduced operator $ \overline{h}_\theta $, which corresponds to $ -\Delta + \Vb $ on $ L^2(S_1) $ with averaged potential (\ref{eq:defVb})
and $ \theta $-boundary
conditions on $ \partial S_1$, can be realized as a Dirichlet form similarly to (\ref{eq:Dirichlet}). One only has to replace $ \po $ by $ \pbo $ in the definition of $ \mathcal{H}_\theta $.
We may thus employ Proposition~\ref{prop-barest}
and the comparison theorem for gaps
\cite[Thm.~1.4]{KirSim87} to estimate
\begin{equation}
    E_0(h_{\theta}) -  E_0(h_{0}) \geq \left(\frac{C_1}{C_2}\right)^2 \left[ E_0\big(\overline{h}_{\theta}\big) -  E_0\big(\overline{h}_{0}\big) \right].
\end{equation}
By the separability of the eigenvalue equation for $ \overline{h}_\theta $ one can compute explicitly
$ E_0\big(\overline{h}_{\theta}\big) -  E_0\big(\overline{h}_{0}\big) = |\theta|^2$, $\theta \in [-\pi, \pi[^{d_1}$,
since $ |\theta|^2 $
is lowest eigenvalue of the Laplacian $ - \Delta $ on $ \mathrm{L}^2(\Lambda_1) $ with
$ \theta $-periodic boundary conditions on $ \partial \Lambda_1 $. This completes the proof of Theorem~\ref{thm:para}. \qed

\section{Proof of Lifshits tails}\label{sec:Lif}

The proof of Theorems~\ref{thm:Lif} and \ref{thm:Lif2} basically follows the lines of reasoning in \cite{KirSim86,Mez87,KiWa}.
We therefore try to be as brief as possible and only focus on the main changes due to the surface nature of the random potential.

The main ingredient of our proof are bounds on the IDSS, which go back to \cite{KirMar83b,Sim85}.
The lower bound employs the reduced-volume IDSS corresponding to Dirichlet
boundary conditions.
The upper bound involves the reduced-volume IDSS corresponding to $ H_{S_L}^\chi(V\ob) $ on
$ L^2(S_L) $ with Mezincescu boundary conditions  (\ref{def:chi}) on $ \partial S_L $.
More precisely, we take the positive ground state $ \psi_0 $ of $ H_{\mathrm{per}} $ for the definition of
(\ref{def:chi}) and impose boundary conditions~(\ref{eq:mezbo}) on functions in the domain of (\ref{eq:schop}).
Both reduced-volume IDSS's are further estimated along the line in \cite{KirMar83a,KirSim86} to yield
\begin{proposition}
Let $ L \in \mathbb{N} $. Under assumptions B1 and B2.1-2.,  S1.1 and S2.1, we have
\begin{align}\label{eq:basin}
    \frac{1}{L^{d_1}} \mathbb{P}\left\{  E_0\big(H_{S_L}^D(V\ob)\big) < E \right\} & \leq N(E) \\
    & \leq \frac{1}{L^{d_1}} \; N\big(H_{\mathrm{per},L}^\chi,E\big) \;\;
    \mathbb{P}\left\{ E_0\big(H_{S_L}^\chi(V\ob)\big) < E \right\} \notag
\end{align}
for all $ E < 0 $.
\end{proposition}

To obtain a lower respectively upper bound on the quantities in (\ref{eq:basin}),
we construct an upper respectively lower bound on
the lowest eigenvalue of $ H_{S_L}^D $ respectively $ H_{S_L}^\chi $.
For this purpose we decompose the given Schr\"odinger operator
\begin{equation}
     H(V\ob) = H_{\mathrm{per}} + V\ob_{\mathrm b} + W\ob
\end{equation}
into the periodic background operator (\ref{eq:background}), the non-negative bulk random potential $ V\ob_{\mathrm b} $
and a non-negative alloy-type surface potential given by
\begin{equation}
W\ob(x) := \sum_{j \in \mathbb{Z}^{d_1}} \rho_j\ob\, f(x_1-j,x_2), \qquad
    \rho_j\ob:=q_j\ob - q_{\mathrm{min}} \end{equation}

\subsection{Upper bound}
The main ingredient herefore is Temple's inequality. Its applicability heavily relies on the lower bound (\ref{eq:gapest}) for
the gap of the periodic background operator $  H_{\mathrm{per},L}^\chi $.

\begin{lemma}\label{lem:temple}
Let $ L \in \mathbb{N} $ and assume there is some function $ 0 \leq  W\ob_R \in L^\infty(\Lambda_L) $ and some
Borel set $ F_2 \subset \mathbb{R}^{d_2} $ such that
\begin{equation}
    W\ob(x) \geq W\ob_R(x_1) \, 1_{F_2}(x_2), \quad\mbox{and}\quad
         \sup_{x_1} \, W\ob_R(x_1) \leq \frac{C_\mathrm{per}}{3 L^2} \label{ass:U2}
\end{equation}
where $C_\mathrm{per} $ is the constant in {\rm (\ref{eq:gapest})}. Then there is some $ C > 0 $ such that
\begin{equation}\label{eq:temple}
     E_0\big(H_{S_L}^\chi(V\ob)\big) \geq E_0 + \frac{C}{L^{d_1}}
    \int_{\Lambda_L} W\ob_R(x_1) \, dx_1.
\end{equation}
\end{lemma}
\begin{proof}
By virtue of B2 and (\ref{ass:U2}) the lowest eigenvalue of $ H_{\mathrm{per}} + W\ob_R 1_{F_2} $ on $ L^2(S_L) $ with Mezincescu boundary
conditions on $ \partial S_L $ is a lower bound on $ E_0\big(H_{S_L}^\chi(V\ob)\big) $. The former may
be lower bounded
with the help of Temple's inequality \cite[Thm.~XIII.5]{ReSi4}.
Choosing $ \psi_0 \in {\mathrm L}^2(S_L) $ as the variational
function, where $ \psi_0 $ is the $ L^2(S_L)$-normalized ground state of
$H_{\mathrm{per}} $,
and using $ H_{\mathrm{per},L}^\chi\psi_0 = E_0 \, \psi_0 $, we obtain
\begin{equation}
  E_0\big(H_{S_L}^\chi(V\ob)\big) \geq
     E_0 + \big\langle \psi_0 ,W\ob_R \, 1_{F_2} \, \psi_0   \big\rangle
    + \frac{\big\langle W\ob_R 1_{F_2}  \psi_0 ,  W\ob_R \, 1_{F_2}\,  \psi_0 \big\rangle }{
        E_1\big(H_{\mathrm{per},L}\big) - E_0 + \langle \psi_0 ,  W\ob_R \, 1_{F_2}\,  \psi_0 \rangle }
\end{equation}
provided the denominator is positive. But this follows from Theorem~\ref{thm:gapest} and assumption~(\ref{ass:U2}) which
imply that the denominator is bounded from below by $ 2 C_\mathrm{per} / 3 L^2 $.
Assumption~(\ref{ass:U2}) also ensures that the numerator is bounded from above by
$ \big\langle \psi_0 ,  W\ob_R 1_{F_2}  \psi_0 \big\rangle $\hspace{0pt}$ C_\mathrm{per}/ 3 L^2 $ so that
\begin{equation}
   E_0\big(H_{S_L}^\chi(V\ob)\big) \geq E_0 +
    \frac{1}{2 L^{d_1}} \int_{S_L} W\ob_R(x_1) \, 1_{F_2}(x_2) \,  \psi_0(x_1,x_2)^2  dx_1 dx_2.
\end{equation}
The proof is completed with the help of the lower bound in (\ref{eq-barest}).
\end{proof}

In order to be able to apply the above lemma, we distinguish to cases:
\begin{itemize}
\item[] {\it Quantum case:} Assumption~S5 is valid.
\item[] {\it Classical case:} Assumption~S5' is valid.
\end{itemize}
In the quantum case we use the fact that there is a constant $ 0< f_u < 1$ and
two Borel sets $ F_1 \subset \Lambda_1$ and
$ F_2 \subset \mathbb{R}^{d_2} $ such that $ f(x) \geq f_u 1_{F_1}(x_1) 1_{F_2}(x_2) $.
Accordingly,
\begin{equation}
    W\ob_R(x_1) := f_u
    \sum_{j \in \mathbb{Z}^{d_1}} \min\left\{ \rho_j\ob , \frac{C_\mathrm{per}}{4L^2}\right\} 1_{F_1}(x_1-j)
\end{equation}
satisfies (\ref{ass:U2}). We may now proceed as in \cite[Prop.~3]{KirSim86} (see also \cite[Lemma~4.4]{KiWa}) and estimate
the integral in (\ref{eq:temple}) in terms of $ f_u $, the Lebesgue measure of $ F_1 $ and
\begin{equation}
   \sum_{ | j |_\infty < \frac{L}{2} } \min\left\{ \rho_j\ob , \frac{C_\mathrm{per}}{4L^2}\right\}
    \geq \frac{C_\mathrm{per}}{4L^2} \; \#\left\{ | j |_\infty < \frac{L}{2} \, \mid \,
        \rho_j\ob \geq  \frac{C_\mathrm{per}}{4L^2} \right\}.
\end{equation}
Choosing $ L $ proportional to $ (E-E_0)^{-1/2} $ with a suitable proportionality constant,
the probability in right-hand side of (\ref{eq:basin})
is therefore estimated from above as follows
\begin{align}
    \mathbb{P}\left\{  E_0\big(H_{S_L}^\chi(V\ob)\big) < E \right\} & \leq
    \mathbb{P}\left\{  \#\left\{ | j |_\infty < \frac{L}{2} \, \mid \,
        \rho_j\ob \geq \frac{C_\mathrm{per}}{4L^2} \right\} < L^{d_1} \right\} \notag \\
    & \leq \exp\left[ - C L^{d_1} \right] \label{eq:upperqm}
\end{align}
where the last inequality stems from the fact that the set is a large deviation event. Reinserting the defining relation of $ L $ to $ E $, we thus obtain an upper bound on $ N $, which proves one of the estimates constituting (\ref{eq:Lif}).\\

In the classical case we pick $ L = 1 $ and set
\begin{equation}
    W\ob_R := f_u \sum_{|j |> R } \min\big\{ \rho_j\ob , 1 \big\} \, \frac{1}{|x_1 -j|^{\alpha}}
\end{equation}
which satisfies (\ref{ass:U2}) by taking $ R > 0 $ small enough. The integral in (\ref{eq:temple}) is then estimated by
$ \int_{\Lambda_L}  W\ob_R(x_1) \, dx_1 \geq C R^{-\alpha} \sum_{R < |j |<2 R} \min\big\{ \rho_j\ob , 1 \big\} $.
Choosing $ R^{\alpha-d_1} $ proportional to $ E-E_0 $ with a suitable
proportionality constant we thus obtain
\begin{align}
    \mathbb{P}\left\{  E_0\big(H_{S_L}^\chi(V\ob)\big) < E \right\} & \leq
    \mathbb{P}\Big\{ R^{-d_1} \!\!\sum_{R < |j |<2 R} \min\big\{ \rho_j\ob , 1 \big\} < c \Big\}
    \notag \\
    & \leq \exp\left[ - C R^{d_1} \right]. \label{eq:uppercl}
\end{align}
Here the last inequality uses the fact that we may choose $ c $ small, such that the last set is a large deviation event.
In total we thus have proved an upper bound on $ N $, which proves one of the estimates constituting (\ref{eq:Lif2}).
\subsection{Lower bound}

The desired upper bound on the lowest eigenvalue of $ H_{S_L}^D(V) $
is basically a consequence of the Rayleigh-Ritz principle and some elementary estimates in \cite{KiWa}.
We summarize these inequalities as
\begin{lemma}
\label{lem:upbouE} Let $ \beta := \max\left\{ 1, \frac{2}{\alpha -d_1}\right\} $.
There are constants $ \gamma $, $ c $, $ C > 0 $ such that
\begin{equation}\label{eq:Ray}
        E_0\big(H_{S_L}^D(V\ob)\big) \leq  E_0 + \sum_{|j|_\infty <L^\beta} \rho_j\ob
    + \frac{C}{L^{d_1}} \int_{\Lambda_{L,M}} \mkern-20mu  V\ob_{\mathrm b}(x)\, dx + e^{-\gamma M} + \frac{c}{L^{2}}
\end{equation}
for all $ L $ and $ M $ large enough.
\end{lemma}
\begin{proof}
 We pick as the variational function the positive, $ \mathbb{Z}^{d_1} $-periodic ground-state
 $ \psi_0 $ of $ H_{\mathrm{per}}$ corresponding to $ E_0 = \infspec H_{\mathrm{per}} $ times a smooth cutoff in order
 to comply with Dirichlet boundary conditions.
 Analogously as in \cite[Lemma~5.1]{KiWa} the Rayleigh-Ritz principle then yields the upper bound
\begin{equation}\label{eq:Rayleigh}
    E_0\big(H_{S_L}^D(V\ob)\big)
    \leq E_0  + C \, L^{-d_1} \int_{S_L} \!\big( W\ob(x) + V\ob_{\mathrm b}(x) \big) \, \psi_0(x)^2 dx + c L^{-2}
\end{equation}
where $ c $, $C > 0 $ are some constants. The $ L^{-2} $-term stems from localizing the function $ \psi_0 $ to $ S_L $.
Using the upper bound in (\ref{eq-barest}), we
estimate the integral
$ \int_{S_L}W\ob(x) \, \psi_0(x)^2 dx \leq C_2^2 \int_{S_L}W\ob(x) \overline{\psi}_0(x_2)^2 dx $.
Denoting $ f_1(x_1) := \int_{\mathbb{R}^{d_2}} f(x_1,x_2) \, \overline{\psi}_0(x_2)^2 \, dx_2$, we may further estimate
\begin{equation} \label{eq:inaus}
    \int_{S_L} \!W\ob(x)\,  \overline{\psi}_0(x_2)^2 dx
    \leq \| f \|_1  \!\!\sum_{|j|_\infty<L^\beta}\!\! \rho_j\ob
    +  | q_{\mathrm{min}}| \!\! \sum_{|j|_\infty \geq L^\beta} \int_{|x_1|_\infty <L}\mkern-30mu f_1(x_1 - j) dx_1.
\end{equation}
Assumption S5 or S5' on $ f $ implies $ f_1(x_1) \leq f_0 \, | x_1 |^{-\max\{d_1 +2 ,\alpha\}} $
such that the inequality \cite[Eq.~(27)]{KiWa}
bounds the second term on the right-hand side of (\ref{eq:inaus}) by a constant times
$ L^{d_1} L^{-\beta \max\{d_1 +2 ,\alpha\}} \leq L^{d_1} L^{-2} $.

To estimate the second part of the integral in (\ref{eq:Rayleigh}) we employ the exponential bound (\ref{eq:expon}), which is
valid for $ \psi_0 $
since Theorem~\ref{prop:exp} covers the case $ V_{\mathrm s} = V_{\mathrm b} = 0 $ and $ L = \infty $.
Picking a cuboid (\ref{eq:defL}) with $ M $ large enough, we obtain
\begin{equation}
    \int_{S_L} \! V\ob_{\mathrm b}(x)\, \psi_0(x)^2 dx
    \leq \| \psi_0 \|_\infty^2  \int_{\Lambda_{L,M}} \mkern-20mu  V\ob_{\mathrm b}(x)\, dx
    + C^2   \int_{S_L \backslash\Lambda_{L,M}} \mkern-30mu V\ob_{\mathrm b}(x)\, e^{-2 \gamma |x_2|} \, dx.
\end{equation}
The last term yields the exponential term in (\ref{eq:Ray}), since $ V\ob_{\mathrm b} \in L^1_{\mathrm{unif}}(\R^d) $.
\end{proof}

We now pick $ M = \frac{2}{\gamma} \ln L $ and $ L $ proportional to $  (E-E_0)^{-1/2} $ and estimate
the probability on the left-hand side in (\ref{eq:basin}) with the help of Lemma~\ref{lem:upbouE}. Choosing the proportionality
constant appropriately this yields
\begin{align}
        &   \mathbb{P}\left\{  E_0\big(H_{S_L}^D(V\ob)\big) < E  \right\} \geq
    \mathbb{P}\Big\{ \sum_{|j|_\infty<L^\beta} \rho_j\ob + \frac{C}{L^{d_1}} \int_{\Lambda_{L,M}} \mkern-20mu  V\ob_{\mathrm b}(x)\, dx
    < \frac{2 C}{L^{2}} \; \Big\}  \notag \\
    &   \qquad    \geq \mathbb{P}_{\mathrm s}\Big\{  \rho_0\ob  \leq \frac{C}{n_0 L^{2+ \beta d_1}}  \Big\}^{n_0 L^{\beta d_1}} \;
    \mathbb{P}_{\mathrm b}\Big\{ \frac{1}{L^{d_1}} \int_{\Lambda_{L,M}} \mkern-20mu  V\ob_{\mathrm b}(x)\, dx
                                       \leq \frac{1}{L^2} \Big\} \notag \\
    & \qquad \geq \left(\frac{C}{ n_0 L^{2+ \beta d_1}}\right)^{\kappa_{\mathrm s } n_0 L^{\beta d_1}}
    \left(\frac{1}{L^2}\right)^{\kappa_{\mathrm s } L^{d_1} M^{d_2}}.
\end{align}
Here the second inequality uses the independence of $ V\ob_{\mathrm b} $ and
the random variables $ (\rho_j)_{j\in \mathbb{Z}^{d_1}} $ and the
fact that there is some constant $ n_0>0 $ such that the number of lattice site $ j \in \mathbb{Z}^{d_1} $ with $ |j|_\infty<L^\beta $ can
be bounded from above by $ n_0  L^{\beta d_1} $. The third inequality rephrases parts of assumptions B2 and S2.
Inserting the defining relation of $ L $ to $ E $ and noting that $ \beta \geq 1 $, we have thus proved that
\begin{equation}
    \liminf_{E \downarrow E_0} \, \frac{\ln \left| \ln N(E)\right| }{\ln( E-E_0)}
      \geq - \frac{\beta \, d_1}{2} = - \max\left\{\frac{d_1}{2}, \frac{d_1}{\alpha - d_1} \right\}
\end{equation}
which together with the lower bounds (\ref{eq:upperqm}) and (\ref{eq:uppercl}) completes
the proof of Theorem~\ref{thm:Lif} and \ref{thm:Lif2}. \qed

\section{Proof of localization}


Localisation for random surface potentials was proved by Boutel de Monvel and Stollmann \cite{BdMS03}
under the assumptions S1-S4 and S6 with $U_b=V_b\equiv0$. Moreover those authors had to impose a
further condition on $P_0$, the distribution of the $q_i$, namely:
\begin{equation}\label{BS}
P_0([q_{min}, q_{min}+\varepsilon])\leq C \varepsilon^{\tau}
\end{equation}
for $\varepsilon > 0$ and some $\tau > 0$.
Boutel de Monvel and Stollmann use (\ref{BS}) as an input for their multiscale analysis. In fact,
(\ref{BS}) allows them to do the initial scale estimate. Those authors remark that a Lifshits tail
estimate would allow them to drop assumption (\ref{BS}).
In this section we briefly explain how one can do the localization proof using the results of the
previous sections  without assuming  (\ref{BS}).
Our multiscale analysis follows the lines of \cite{BdMS03,Sto01,KSS98b,KSS98a}.
The reader is referred to those works for details. We will mainly sketch a few differences.

It is customary to use (bounded) cubes $[-\frac{L}{2}, \frac{L}{2}]^d$ as building blocks for
multiscale analysis. However, to emphasize the underlying geometry of our problem, we suggest to use strips
$S_L:=[-\frac{L}{2}, \frac{L}{2}]^{d_1}\times\R^{d_2}$ instead.

For simplicity, we suppose in what follows that the bulk random spectrum $V_b$ vanishes. With a
little more effort one could handle the general case using the same technique.

One of the key ingredients of multiscale analysis is a form of the Wegner estimate which we prove
as in [BdMS03].

\begin{proposition}
For $E>0$ and $\varepsilon >0$ we have

\begin{equation}
\mathbb{P}\left(\sigma(H^D_{S_L})\ \cap\ ]E-\varepsilon, E+\varepsilon[\ =\emptyset\right)\leq
CL^{2d_1}\varepsilon^{\mu}.
\end{equation}
{\rm [}Here the exponent $\mu$ was defined in S6 and the constant $C$ may depend on $E$, but not on $L$ or
$\varepsilon$.{\rm ]}
\end{proposition}

For a proof we refer to \cite{BdMS03} and \cite{Sto00}. Our Wegner estimate above has an
upper bound of order $L^{2d_1}$ which suffices to prove Anderson localization. However,
to prove H\"older continuity of the integrated density of surface states we would need
a bound proportional to $L^{d_1}$. We believe that such a bound can be done using more
elaborated techniques as in \cite{CHN01,CHKN02,KoSch01} for example.

The second input to multiscale analysis is an initial scale estimate.
To do the initial scale estimate Boutel de Monvel and Stollmann use the additional assumption
~(\ref{BS}). We base our initial scale estimate on the Lifshitz tails result of the previous chapter,
thus avoiding any additional assumption like ~(\ref{BS}).
\begin{proposition}
There exists $E_1 >E_0$, and $L_1\in \mathbb{N} $ and a constant $C$, such that
  \begin{equation}\label{eq:PL}
    \mathbb{P}\left\{ E_0\big(H^D_{S_L}(V\ob)\big)<E\right\} \leq L^{d_1}  e^{-C(E-E_0)^{d_1/4}}
  \end{equation}
for all  $L\geq L_1$ and $E\leq E_1$ we have.
\end{proposition}
\begin{proof}
    This follows directly from the lower bound in (\ref{eq:basin}) together with the Lifshits asymptotics (\ref{eq:Lif})
    in Theorem~\ref{thm:Lif}.
\end{proof}
The induction step of the multiscale analysis is done along the usual lines
(see \cite{BdMS03} and \cite{Sto01} for details).

\section*{Acknowledgment}
This work was supported by the DFG within the SFB TR 12 and grant no.\ Wa 1999/1-1.


\end{document}